\documentclass[11pt]{article} 
\usepackage{graphicx}

\usepackage{times}
\usepackage{listings}
\usepackage{amssymb}

\title{A New Method for Avoiding Data Disclosure While
Automatically Preserving Multivariate Relations}
\author{
Norman Matloff\thanks{Dept. of Computer Science, University of
California, Davis}
\and
Patrick Tendick\thanks{Avaya Corp.}
}

\begin{document}

\maketitle

\begin{abstract}

\noindent Statistical disclosure limitation (SDL) methods aim to provide
analysts general access to a data set while limiting the risk of
disclosure of individual records.  Many methods in the existing
literature are aimed only at the case of univariate distributions, but
the multivariate case is crucial, since most statistical analyses are
multivariate in nature.  Yet preserving the multivariate structure of
the data can be challenging, especially when both continuous and
categorical variables are present.  Here we present a new SDL method
that automatically attains the correct multivariate structure,
regardless of whether the data are continuous, categorical or mixed,
and without requiring the database administrator to estimate that
multivariate structure.  In addition, operational methods for assessing
data quality and risk will be explored.

\end{abstract}

\section{Introduction}

Statistical disclosure limitation (SDL) methods aim to provide analysts
statistical access to a data set while limiting the risk of disclosure of
individual records.  Common methods include noise addition, swapping of
parts of records, replacing data by synthetic equivalents, suppression
of small cells in contingency tables, and so on \cite{duncan}. 


Long a field of statistical research, in recent years SDL issues have
attracted the interest of computer scientists \cite{dwork}.  There has
been a marked contrast in the approaches taken by the two communities:
The statistical view is that of serving research analysts who wish to do
classical inference on samples from populations, while the computer
scientists, coming from a cryptographic background, have viewed the data
itself as the primary focus.  In other words, in the computer science
approach, the `S' in SDL has perhaps had lesser attention, compared to
the statisticians' view of things.  However, there is some indication of
increasing interaction between the two groups \cite{abowd}.

For an overview of how methodology has been refined and expanded over
time, compare a 1989 survey paper \cite{adam}, a 2002 Census Bureau
viewpoint \cite{census2002}, the current statistical view \cite{duncan},
and the more recent computer science approach \cite{dwork}.

Whatever approach is taken, a primary goal remains statistical analysis
by the end user.  And in order to perform meaningful statistical
analysis on the data, {\bf one's methods must at least approximately
preserve multivariate structure}.  Most ststistical analysis --- linear
regression, logistic models, princple components analysis, the
log-linear model and so on -- are inherently multivariate.
Unfortunately, many existing SDL methods place little or no emphasis on
this aspect, and this is an absolutely central issue.  Regression
coefficient estimates, for instance, can turn out substantially biased
as a result.  As noted in \cite{nivule},

\begin{quote}
...[in using] noise addition techniques...the original data
suffers loss of some of its statistical properties even while
confidentiality is granted, thus making the dataset almost meaningless
to the user of the published dataset. 
\end{quote}

The above statement applies only to independent noise variables. Noise
addition methods can preserve the multivariate structure of continuous
variables, if the data come from an approximate multivariate normal
distribution, by adding correlated noise \cite{matloff1986}
\cite{kim} \cite{tendick}.  However, this does not apply to the
discrete-variable case, and moreover, the same problems apply to most if
not all of the other major classes of SDL methods.

Developing methodology for the mixed continuous/discrete case is a
difficult problem; see \cite{manrique} and the citations therein for
some existing methodology.  To broaden the methods available to Data
Stewardship Organizations (DSOs), a new method is proposed in this paper
to deal with the multivariate structure preservation problem. Our method
has several important advantages:

\begin{itemize}

\item The method works on general data, i.e.\ continuous, discrete or
mixed. 

\item The method does not require the DSO to estimate the dependency
structure between the variables, or make assumptions regarding that
structure.

\item The method has several tuning parameters, affording DSO
broad flexibility in attaining the desired balance between
privacy and statistical usability.

\end{itemize}

\section{Overview of the Method}
\label{overview}

Let $W_{ij}, i = 1,...n, j = 1,...,p$ denote our original data on $n$
individuals and $p$ variables.  Choose $\epsilon > 0$ and $0 < q \leq
1$.  Then we form our released data $W'_{ij}$ as follows:

For $i = 1,...n$:

\begin{itemize}

\item Consider record $i$ in the data base:

\begin{equation}
r_i = (W_{i1},...,W_{ip})
\end{equation}

\item With probability $1-q$, skip the next steps.

\item Find the set $S$ of points in the data set within $\epsilon$
distance of (but excluding) $r_i$.

\item Draw a random sample (\underline{with} replacement) of $p$ items
from $S$, resulting in values $a_{km}, k = 1,...,p, m = 1,...,p$.

\item For $j = 1,...,p$, set 

\begin{equation}
W'_{ij} = a_{jj}
\end{equation}

and store the released, modified version of $r_i$ as

\begin{equation}
r_i' = 
(W_{i1}',...,W_{ip}')
\end{equation}

\end{itemize}

\section{Theoretical Justification}

Note carefully that the procedure described in the last section {\it
does not rely on knowledge or estimation of the multivariate
distribution of our data}, a key advantage of the methodology we are
proposing here.  On the contrary, the components of $r_i'$ are generated
independently.  The following result shows that the multivariate
structure is (approximately) preserved anyway.  For expositional
convenience, the theorem and proof will be stated for the case $p = 1$.

{\bf Theorem:}  Consider a bivariate random vector $(X,Y)$ and $\epsilon
> 0$.  For any $t$ in $\mathcal{R}^2$, let $A_{t,\epsilon}$ denote the
$\epsilon$ neighborhood of $t$, defined by some metric $\mathcal{M}$.
Let $F$ denote the cdf of $(X,Y)$, and define $G_{t,\epsilon}$ to be the
conditional cdf of $(X,Y)$, given that that vector is in
$A_{t,\epsilon}$.  Finally, given $(X,Y)$, define
\underline{independent} random variables $U$ and $V$ to be drawn
randomly from the first- and second-coordinate marginal distributions of
$G_{(X,Y),\epsilon}$, respectively.  Then

\begin{equation}
\lim_{\epsilon \rightarrow 0}
P
\left (
U \leq a \textrm{ and } V \leq b
\right )
= F(a,b)
\end{equation}

for all $-\infty < a,b < \infty$.  

In other words, as $\epsilon$ goes to 0, the unconditional bivariate
distribution of $(U,V)$ goes to that of $(X,Y)$, {\it even though $U$
and $V$ are conditionally independent}.

{\bf Proof:}

Given $(X,Y) = t = (t_1,t_2)$,

\begin{equation}
\lim_{\epsilon \rightarrow 0} U = t_1
\end{equation}

and

\begin{equation}
\lim_{\epsilon \rightarrow 0} V = t_2
\end{equation}

Then by bounded convergence,

\begin{eqnarray}
\lim_{\epsilon \rightarrow 0}
P
\left (
U \leq a \textrm{ and } V \leq b
\right )
&=& 
\lim_{\epsilon \rightarrow 0}
E \left [
P
\left (
U \leq a \textrm{ and } V \leq b
~|~ X,Y \right )
\right ] \\ 
&=& 
\lim_{\epsilon \rightarrow 0}
E \left [
P(U \leq a ~|~ X,Y ) \cdot
P(V \leq b ~|~ X,Y ) 
\right ] \\
&=& E \left [
1_{X \leq a} \cdot
1_{Y \leq b}
\right ] \\
&=& E \left [
1_{X \leq a \textrm{ and } Y \leq b}
\right ] \\
&=& P(X \leq a \textrm{ and } Y \leq b) \\
&=& F(a,b)
\end{eqnarray}

$\blacksquare$

\bigskip

The key word {\it independent} in the above theorem has a major
implication:  We can make our released data approximate the multivariate
distribution of the original data (or the population from which the
latter are drawn), {\bf without knowing or even estimating the
multivariate relationship of our variables}.  We simply sample {\it
independently} from $S$, yet attain the correct {\it dependency}
relationship among the variables.





\section{Code and Tuning Parameters}

The method provides the DSO with excellent flexibility in achieving the
desired balance between privacy and accurate multivariate structure, via
the following tuning tuning parameters:

\begin{itemize}

\item The neighborhood radius, $\epsilon$.

\item The distance metric $\mathcal{M}$.

\item The proportion $q$ of modified records.  

\end{itemize}

Code implementing the method is provided on GitHub, at {\it
https://github.com/matloff/statdb}.\footnote{Publicly available software 
for existing SDL methods includes sdcMicro on CRAN and WebSwap at NISS. }  
The call form is

\begin{lstlisting}
nbrs(z, eps, modprop = 1, wts = NULL) 
\end{lstlisting}

where {\bf eps} is $\epsilon$, {\bf modprop} is $q$, and the {\bf wts}
argument exerts some control on the distance metric, to be explained
shortly.  The return value is the released data set, in the form of an R
data frame (which could be converted to SQL etc.).

It is assumed that all categorical variables have been converted to
dummy variables.  Ordinary Euclidean distance is used on the scaled
data, including any dummy variables.  Scaling places all the variables
on the same footing --- all now have standard deviation 1 --- but there
is still a difference between the continuous variables and the dummies
and other discrete variables, as follows.

As sample size $n$ grows (treating the original data as a sample from
some population), one would want $\epsilon$ to become smaller, but this
would not work well for the discrete variables.  With large $n$, the
latter would come to dominate the distance metric, and one could not
drop $\epsilon$ below some minimum threshhold.  The {\bf wts}
argument provides the DSO with a tool to reduce that dominance, by
allowing the weights of the discrete variables (or others) to decrease
as $n$ increases.

If for example we set {\bf wts = c(5,12,13,rep(0.6,3))}. then in
computing distances the variables in columns 5, 12 and 13 of the data
matrix are reduced in weight by a factor of 0.6.  

\section{Selection of Tuning Parameters}

In some modern statistical methods, the user is faced with selection of
a large number of tuning parameters, both numeric and policy-oriented,
such as in the SIS package \cite{fan}.  The user may find the task of
setting those parameters daunting and bewildering.

In SDL settings, though, the DSO may {\it welcome} the selection of
tuning parameters. The goal is achieving a good balance between
statistical accuracy of the released data and disclosure risk, a
difficult task, so from the DSO's point of view, the more tuning
parameters the better.

\subsection{Choices}

For a given set of tuning parameters, the DSO wishes to assess

\begin{itemize}

\item [(a)] whether the results of statistical  analyses on the released
data set will be reasonbly close to those of the original data, and 

\item [(b)]
whether records that were at risk in the original data will be masked 
sufficiently well in the released data.

\end{itemize}

For both (a) and (b), we propose an operational approach.\footnote{We
have not seen this in the literature, though it is likely that some DSOs
have experimented with this approach.}  For (a), though many authors have
proposed global measures of distance between the original and released
data sets, we suggest gauging the statistical accuracy of the latter in
a more direct manner, motivated by the intended usage of the data,
namely statistical analyses.

In other words, under this approach the DSO would run several
representative statistical analyses, say regression and principle
components analysis (PCA), on both the original and released data sets.
The DSO would then compare the results. 

Our approach to issue (b) is similarly practical.  The DSO identifies
some representative unique or rare records, and then tracks what happens
to them in the released data.  Have they been hidden sufficiently well?

We advocate these methods (which of course can be used in conjunction
with other methods) because they expose the system in ways that {\it
directly} address the goals (a) and (b):

\begin{itemize}

\item No matter what SDL method is used -- noise addition, cell
suppression, data swapping, our method introduced in this paper, etc.
--- it will necessarily result in some distortion to statistical analyses.
The fact that two (empirical) distributions are close  of course does
not imply that a given functional will have similar values on those two
distributions.

Thus is vital to get a {\it direct} idea of how much distortion the
statistical users of the data may need to tolerate.  This is what our
approach addresses.

\item An example in some of the SDL literature has involved preserving
the privacy of the lone female electrical engineer in a company
employee database.  The DSO can pose questions like this for their given
data set, and find that, say,  while the female EE was hidden, the lone
programmer over age 50 was not, and then continue to search for good
combinations of the tuning parameters..

\end{itemize}

\subsection{The Roles of $n$ and $p$}

In setting these parameters, the DSO must take into account not only the
desired balance between (a) and (b) above, but also the values of $n$
and $p$.  For fixed $p$, the larger $n$ is, the fewer the number of
uniquely identifiable individuals in the data, and thus the decreased
need for privacy actions.\footnote{As noted, we are treating the data
as a sample from some (tangible or conceptual) population.  As such, the
notion of a {\it population unique}, seen in some of the SDL literature,
doesn't apply.  If a combination of the categorical variables appears in
our data, then by definition that combination has nonzero probability in
the population, and we'll get more and more individuals of that type as
$n$ grows.  For continuous variables, a similar statement holds in the
sense that as $n$ grows, we will have more and more individuals near the
given value.} On the other hand, for fixed $n$, the larger the value of
$p$, the more potential identifiable uniques.

\section{Example}

We used the Census data set in the package {\bf regtools} ({\it
https://github.com/matloff/regtools}) to simulate an employee
database, sampling 5000 records from this data.\footnote{Since this is
just an illustration, the data were not cleaned, and some WageIncome
values were 0 that probably should have been designated as missing.}


The call used was

\begin{lstlisting}
> p1p <- nbrs(p1,eps=0.3,wts=c(2,4,5,rep(0.2,3)))
\end{lstlisting}

To gauge how close this new version of the data was to the original, we
ran a linear regression analysis, predicting WageIncome from Age,
Gender, WeeksWorked, MSDegree and PhD.  The estimated coefficients for
the original and modified data were

\begin{tabular}{|r|r|r|r|r|r|}
\hline
data & Age & Gender & WeeksWorked & MS & PhD \\ \hline 
original & 447.2 & -9591.7 & 1286.4 & 17333.0 & 21291.3 \\ \hline 
released & 466.1 & -8423.2 & 1270.7 & 18593.9 & 22161.4 \\ \hline 
\end{tabular}

The results are fairly good, differing between 1\% and 12\% from the
original.  And the differences are not bad when viewed in the context of
the standard errors of the original:

\begin{tabular}{|r|r|r|r|r|}
\hline
Age & Gender & WeeksWorked & MS & PhD \\ \hline 
52.8 & 1301.9 & 38.6 & 1453.7 & 3627.7 \\ \hline
\end{tabular}

Presumably we could do better with other values of the tuning
parameters.  But what about disclosure risk?

In the original data set, there was one female worker with age under 31:

\begin{lstlisting}
> p1[p1$sex==2 & p1$phd==1 & p1$age < 31,]
          age sex wkswrkd ms phd wageinc
7997 30.79517   2      52  0   1  100000
\end{lstlisting}

How well was she hidden in the modified data?  Quite well, it turns out:

\begin{lstlisting}
> p1pc <- na.omit(p1p)
> p1pc[p1pc$sex==2 & p1pc$phd==1 & p1pc$age < 31,]
           age sex wkswrkd ms phd wageinc
12522 30.5725   2      52  0   1   50000 
\end{lstlisting}

There is one person listed in the released data of the given description
(female, PhD, age $<$ 31).  But she is listed as having an income of
\$50,000 rather than \$100,000.  In fact, it is a different person, worker
number 12522, not 7997.\footnote{Of course, ID numbers would be suppressed.}
Where is the latter now?

\begin{lstlisting}
> which(rownames(p1p) == 7997)
[1] 3236
> p1p[3236,]
         age sex wkswrkd ms phd wageinc
7997 31.9746   1      52  0   1  100000
\end{lstlisting}

Ah, she became a man!  That certainly hides her.  

This is just a first try.  The DSO could continue, experimenting with
various other values of the tuning parameters.  For instance,
we tried raising the weight of the categorical variables:

\begin{lstlisting}
> p1p <- nbrs(p1,eps=0.6,wts=c(2,4,5,rep(0.3,3)))
\end{lstlisting}

The new regression coefficients were generally good:

\begin{tabular}{|r|r|r|r|r|r|}
\hline
data & Age & Gender & WeeksWorked & MS & PhD \\ \hline 
relased & 506.3  & -9323.1  & 1289.8 & 17684.1 & 22019.3 \\ \hline 
\end{tabular}

Now there were no workers in the modified data set satisfying
the given conditions: 

\begin{lstlisting}
> p1pc <- na.omit(p1p) 
> p1pc[p1pc$sex==2 & p1pc$phd==1 & p1pc$age < 31,]
[1] age     sex     wkswrkd ms      phd     wageinc
<0 rows> (or 0-length row.names)
\end{lstlisting}

What happened was that worker 7997? She had no close neighbors other than
herself, so her data became NAs:

\begin{lstlisting}
> p1p[3236,]
     age sex wkswrkd ms phd wageinc
7997  NA  NA      NA NA  NA      NA
\end{lstlisting}

So again this worker 7997 was protected.

This of course just begins to explore the various tuning
parameter values that the DSO could experiment with, in addition to
doing so on other types of analyses, say principle components analysis.



\section{Other Types of Privacy}

Another type of privacy may need to be considered.  Think of our example
of the lone female electrical engineer in an employee database.  Our
concern there is that an intruder may know that she is in the database,
and may know enough identifying information about her that he may be
able to determine which record is hers, and thus gain access to
sensitive information.  But in some cases mere knowledge that a given
individual is actually in the database can itself be sensitive
information.

For instance, consider a cancer patient who wishes to participate in a
clinical trial, but is concerned that his diseased status may become
public knowledge.  Suppose further that an intruder knows that this
patient was born in Tonga, and that the intruder is fairly sure that
there is only person in the community with that characteristic.  Our
proposed method may result in some record in the released data showing a
birthplace of Tonga, in which case the nefarious user knows that the
patient does have the disease --- even if the record in the released
data is not for the original patient.  In such a situation, the DSO may
consider excluding this person from the database, or adjusting some of
the tuning parameters.

\section{Discussion and Future Work}

We have proposed a new SDL method that works for mixed
continuous/categorical data and does not require estimation of
multivariate structure.  Our brief preliminary exploration seems
promising.  Much more investigation needs to be done, with different
data sets and more thorough search for good combinations of tuning
parameters.

Note that ``a little bit of privacy can go a long way'':  As long as the
intruder knows that the data have been modified (even for the
nonsensitive variables), there may be enough doubt in his/her mind as to
make the data useless for nefarious purposes (while still being very
useful for legitimate purposes).  Thus, values less than 1.0 for $q$,
the proportion of modified records, will be feasible in some settings.
Perhaps a taxonomy of such settings could be developed.

In databases with large $p$, one must take into account the Curse of
Dimensionality \cite{beyer}.  The DSO may choose to use a weighted
distance metric, with the weights going to 0 as the variable index goes
to infinity \cite{matloff2015}.

In general, the choice of $\epsilon$ must also be made carefully This
approach does require fairly large data sets, so that for instance the
set $S$ contains some female workers in our examples above.  One avenue
of future research would be to investigate allowing the value of
$\epsilon$ to vary from record to record.  

Another point to be investigated concerns records on the fringes of the
data, say far from the centroid under our metric $\mathcal{M}$.  For
such a record, the neighborhood will likely be empty unless we make
$\epsilon$ large, which would create its own problems in terms of
statistical accuracy; observations on the fringes of a data set tend to
have high leverage.  Alternatively, we could use the k-nearest neighbor
method to form our neighborhoods, guaranteeing that they will be
nonempty, but our neighborhoods may again be very large for records on
the fringes.

Accordingly, one aspect of future work will involve the efficacy of
encouraging users of the released data to use outlier-robust methods,
such as robust regression and robust PCA.

{}

\end{document}